\documentclass[manuscript]{aastex61}
\usepackage{graphicx,epstopdf,epsfig}
\usepackage{natbib}
\usepackage{amsmath,amstext}
\usepackage{multirow}
\usepackage{color}
\shortauthors{Dash et al.}

\begin{document}

\title{The 2019 July 2 Total Solar Eclipse: Prediction of the Coronal Magnetic Field Structure and Polarization Characteristics}

\author{Soumyaranjan Dash}
\affiliation{Center of Excellence in Space Sciences India, Indian Institute of Science Education and Research Kolkata, Mohanpur 741246, West Bengal, India}

\author{Prantika Bhowmik}
\affiliation{Center of Excellence in Space Sciences India, Indian Institute of Science Education and Research Kolkata, Mohanpur 741246, West Bengal, India}

\author{Athira B S}
\affiliation{Center of Excellence in Space Sciences India, Indian Institute of Science Education and Research Kolkata, Mohanpur 741246, West Bengal, India}

\author{Nirmalya Ghosh}
\affiliation{Center of Excellence in Space Sciences India, Indian Institute of Science Education and Research Kolkata, Mohanpur 741246, West Bengal, India}
\affiliation{Department of Physical Sciences, Indian Institute of Science Education and Research Kolkata, Mohanpur 741246, West Bengal, India}

\author{Dibyendu Nandy}
\correspondingauthor{Dibyendu Nandy}
\email{dnandi@iiserkol.ac.in}
\affiliation{Center of Excellence in Space Sciences India, Indian Institute of Science Education and Research Kolkata, Mohanpur 741246, West Bengal, India}
\affiliation{Department of Physical Sciences, Indian Institute of Science Education and Research Kolkata, Mohanpur 741246, West Bengal, India}


\begin{abstract}

On 2019 July 2 a total solar eclipse -- visible across some parts of the Southern Pacific Ocean, Chile and Argentina -- will enable observations of the Sun's large-scale coronal structure. The structure of the Sun's corona and their emission characteristics are determined by underlying magnetic fields which also govern coronal heating and solar eruptive events. However, coronal magnetic field measurements remain an outstanding challenge. Computational models of coronal magnetic fields serve an important purpose in this context. Earlier work has demonstrated that the large-scale coronal field is governed by slow surface flux evolution and memory build-up which allows for prediction of the coronal structure on solar rotational timescales. Utilizing this idea and based upon a 51 day forward run of a data-driven solar surface flux transport model and a Potential Field Source Surface model, we predict the Sun's coronal structure for the 2019 July 2 solar eclipse. We also forward model the polarization characteristics of the coronal emission from the predicted magnetic fields. We predict two large-scale streamer structures and their locations on the east and west limbs of the Sun and discuss the possibility of development of a pseudo-streamer based on an analysis of field line topology. This study is relevant for coronal magnetometry initiatives from ground-based facilities such as the Daniel K. Inouye Solar Telescope and Coronal Multichannel Polarimeter, and upcoming space-based instruments such as the Solar Ultraviolet Imaging Telescope and the Variable Emission Line Coronagraph on board ISRO's Aditya-L1 space mission.

\end{abstract}

\section{Introduction}

The Sun's surface and its outer atmosphere -- the corona -- forms the time-dependent inner boundary of the heliosphere, and therefore, forces it. Magnetic field dynamics in the corona including magnetic reconnection and heating is induced by photospheric flux emergence and evolution. In turn this coronal dynamics spawn the solar wind, solar flares, coronal mass ejections (CMEs) and energetic particle flux and high energy radiation which collectively create hazardous space environmental conditions. Understanding the origin of space weather at the Sun and building predictive capabilities is a high priority goal in the space sciences \citep{Schrijver2015AdSpR}. Although the origin of space weather can be traced back to the dynamics of magnetic fields in the Sun's corona, direct observations of coronal magnetic fields is challenging because of the low photon flux associated with the tenuous coronal plasma. Thus computational approaches based on our theoretical understanding of how the Sun's magnetic fields emerge through the surface, evolve and permeate the corona are essential to model and understand coronal dynamics. 

The Alfv\'{e}n speed in the corona (the speed at which magnetic disturbances propagate) is much faster (of the order of 1000 kms$^{-1}$) than the large-scale surface motions that drive coronal evolution ($1$ to $3$ kms$^{-1}$). Consequently, a reasonable approach in theoretical models is to assume that the coronal magnetic field distribution evolves quasi-statically in response to photospheric forcing. These models can be categorized into four broad classes: potential field source surface extrapolation models, force-free models, magnetohydrostatic models, and full magnetohydrodynamic (MHD) models. While the first three primarily provide the three-dimensional close-to-equilibrium magnetic field structure in the solar corona, the last approach can self-consistently provide the magnetic field as well as thermal characteristics. All these modeling approaches have their own advantages and limitations and often a particular model is chosen based on the balance of convenience and sophistication necessary for a particular problem. For a detailed account on coronal magnetic field models see \cite{Mackay2012LRSP,Yeates2015SoPh}.
 
Testing these theoretical modeling approaches through ``true'' predictions and constraining them through observations is best achieved at the time of solar eclipses when lunar occultation masks the bright solar disk revealing the Sun's faint coronal structure. In an earlier work, \cite{Nandy2018ApJ} postulated a novel methodology for long-term coronal field prediction. Post-eclipse assessment demonstrated that they successfully predicted all the major large-scale structures of the Great American solar eclipse (2019 August 21) including a pseudo-streamer that was not captured in an alternative prediction methodology \citep{Mikic2018NatAs}. 
 
\cite{Nandy2018ApJ} used a data-driven Predictive Solar Surface Flux Transport (PSSFT) model developed at CESSI \citep{Bhowmik2018Natcomm} to first predict the surface magnetic field distribution and then feed this into a potential field extrapolation model to obtain the prediction of the coronal magnetic field structure. The PSSFT model simulates the evolution of the surface magnetic field driven by the emergence of tilted active regions and redistribution of associated magnetic flux mediated via large-scale plasma flows and supergranular diffusion. This process is known as the Babcock Leighton (B-L) mechanism \citep{Babcock1961ApJ,Leighton1969ApJL,Wang1989Sci,Ballegooijen1998ApJ,Schrijver2001ApJ,Sheely2005LRSP,Cameron2010ApJ,Mackay2012LRSP} in the context of dynamo theory and bridges solar inetrnal dynamics with coronal field evolution. 

The CESSI PSSFT model used in \cite{Nandy2018ApJ} is driven by observed sunspot data over century time scale and calibrated with polar flux observations spanning multiple solar cycles \citep{Bhowmik2018Natcomm,Munoz2012ApJ}. The slow evolution of the photospheric magnetic field aids in building up a long-term memory (due to mostly deterministic surface flux transport processes) in the PSSFT model. This enables the model to make long-term predictions of the large-scale surface field distribution including the high latitude polar fields \citep{Bhowmik2018Natcomm} -- which can be used as inputs for making coronal field predictions. An accurately predicted surface magnetic field map ensures evaluation of large-scale structures of the coronal magnetic field \citep{Schrijver2003SoPh} with better precision. Under low plasma-$\beta$ (i.e., magnetic pressure dominating over gas pressure) and current-free approximations the potential field source surface (PFSS) extrapolation technique \citep{Altschuler1969SoPh,Schatten1969SoPh} may be employed to simulate coronal magnetic fields. We note that dynamic coronal simulations based on magnetofrictional approaches transmit this surface memory to the corona \citep{yang1986apj,ballegooijen2000apj,yeates2014solarphy}, while PFSS-like extrapolation relies on the memory of the surface field itself.

Here we utilize the \cite{Nandy2018ApJ} prediction methodology -- a combination of the PSSFT and PFSS models -- to predict the large-scale coronal magnetic structure of the 2019 July 2 total solar eclipse. A short research note with the predicted coronal magnetic structure is communicated in \cite{Dash2019rnaas}. This work contains description of the methodology with detailed in-depth analysis of the coronal magnetic field structure and topology including the possibility of a pseudo-streamer appearing in the future. Additionally, here we present forward-modeled polarization maps for the coronal emission expected during the eclipse which may be compared with coronal magnetometry observations. The rest of the paper is structured as follows: a brief description of the computational models is provided in section~2; results are presented in Section~3; concluding discussions follow in Section~4.

\section{Numerical Models and Sunspot Data Input}

Evolution of magnetic field on the solar surface is governed by the magnetic induction equation. As photospheric magnetic field is predominantly in the radial direction \citep{Solanki1993SSRv}, in our PSSFT model we solve only the radial component ($B_r$) of the induction equation which in spherical polar coordinates is given by 
\begin{equation}
\frac{\partial B_r}{\partial t} = -  \omega(\theta)\frac{\partial B_r}{\partial \phi} - \frac{1}{R_\odot \sin \theta} \frac{\partial}{\partial \theta}\bigg(v(\theta)B_r \sin \theta \bigg)
+\frac{\eta_h}{R_\odot^2}\bigg[\frac{1}{\sin \theta} \frac{\partial}{\partial \theta}\bigg(\sin \theta \frac{\partial B_r}{\partial \theta}\bigg) + \frac{1}{\sin \theta ^2}\frac{\partial ^ 2 B_r}{\partial \phi ^2}\bigg] + S(\theta, \phi, t).
\label{ind_surf}
\end{equation}

\noindent The symbols, $\theta$ and $\phi$ represent co-latitude and longitude, $R_\odot$ is the solar radius. The large-scale velocity fields, the differential rotation and meridional circulation on the solar surface are denoted by $\omega(\theta)$ and $v(\theta)$, respectively. These plasma flows are modeled using empirical functions \citep{Bhowmik2018Natcomm}, which are observationally verified \citep{Snodgrass1983ApJ,Ballegooijen1998ApJ}. The parameter $\eta_h$ is the effective diffusion coefficient associated with the turbulent motion of supergranules and $S(\theta, \phi,t)$ is the source term describing the emergence of new sunspots. We note that any new spot is included in the PSSFT model when it has the maximum area coverage on the solar surface during its lifetime (or period of visibility on the solar disk).

The century-scale PSSFT simulation utilizes sunspot data recorded by Royal Greenwich Observatory (RGO) and United States Air Force (USAF)/National Oceanic and Atmospheric Administration (NOAA) for the period 1913 August until 2016 September \citep{Bhowmik2018Natcomm}. The database provides essential details associated with active regions such as position on the solar surface, area coverage, and the corresponding time, etc. From 2016 October onwards, the observed sunspot data used in the PSSFT model is acquired from the Helioseismic and Magnetic Imager (HMI) on board NASA's Solar Dynamic Observatory (SDO). The sunspot area given by HMI is scaled down by a constant factor for consistency with the RGO-NOAA/USAF recorded area; this cross-instrument calibration is necessary as the original calibration of the century-scale PSSFT simulation is based on the later database. The requirement of such cross-instrumental calibration arises due to the different techniques employed for recording sunspot data \citep{Munoz2015ApJ}. The scaling factor is evaluated by comparing the area associated with 1174 active regions recorded in both the NOAA/USAF and HMI database for an overlapping period of six years (2010 May -- 2016 September). We measure the ratio between the area recorded by the two databases and use a Gaussian fit to the distribution for calculating the mean. The mean ratio ($0.244$) is utilized to scale down the HMI-recorded area. The magnetic flux of the active region is evaluated based on a linear empirical relation \citep{Sheeley1966ApJ,Dikpati2006GeoRL}: $\Phi(A) = 7.0 \times 10^{19} A$ maxwells, where $A$ is the area in unit of micro-hemispheres. We assume all active regions appearing on the solar surface are ideal $\beta$-spots where the magnetic flux is equally distributed between the leading and following polarity spots. The associated tilt angle is decided by Joy's law with a cycle-dependent refinement (see equation 5 of \citealt{Bhowmik2018Natcomm}). 

The last observed active region included in the PSSFT model is AR12741 -- which attained its maximum area coverage on 2019 May 12. The PSSFT model is then forward run until the day of the eclipse on 2019 July 2 (assuming no new sunspot emergence till then) to generate the predicted surface magnetic field distribution.

To model the corresponding coronal magnetic field structure we use an PFSS extrapolation technique with the predicted surface magnetic field used as the bottom boundary of the computational domain. The extrapolation is extended up to the source surface ($2.5 R_{\odot}$) beyond which we assume the magnetic field to become radial \citep{Davis1965IAUS}. We utilize the PFSS extrapolation model developed by \cite{anthony_yeates_2018_1472183}\footnote{{\href{https://github.com/antyeates1983/pfss}{\color{black}Github link for the code: \color{blue}https://github.com/antyeates1983/pfss}}}.

\section{Results}

\subsection{Prediction of the Coronal Magnetic Field Structure}

\begin{figure}[!htb]
\centering
\includegraphics[width=0.80\linewidth]{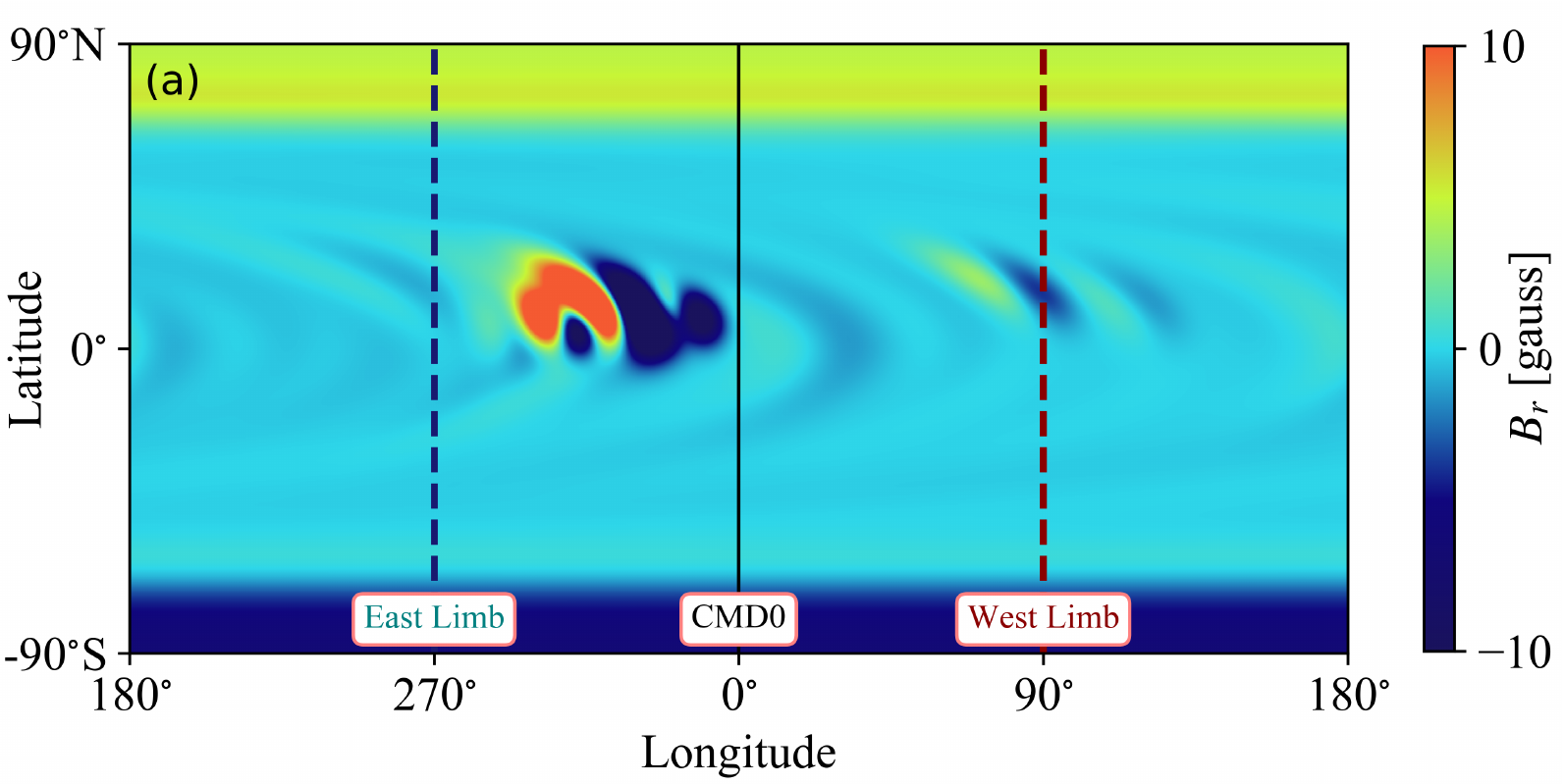}\\
\includegraphics[width=0.80\linewidth]{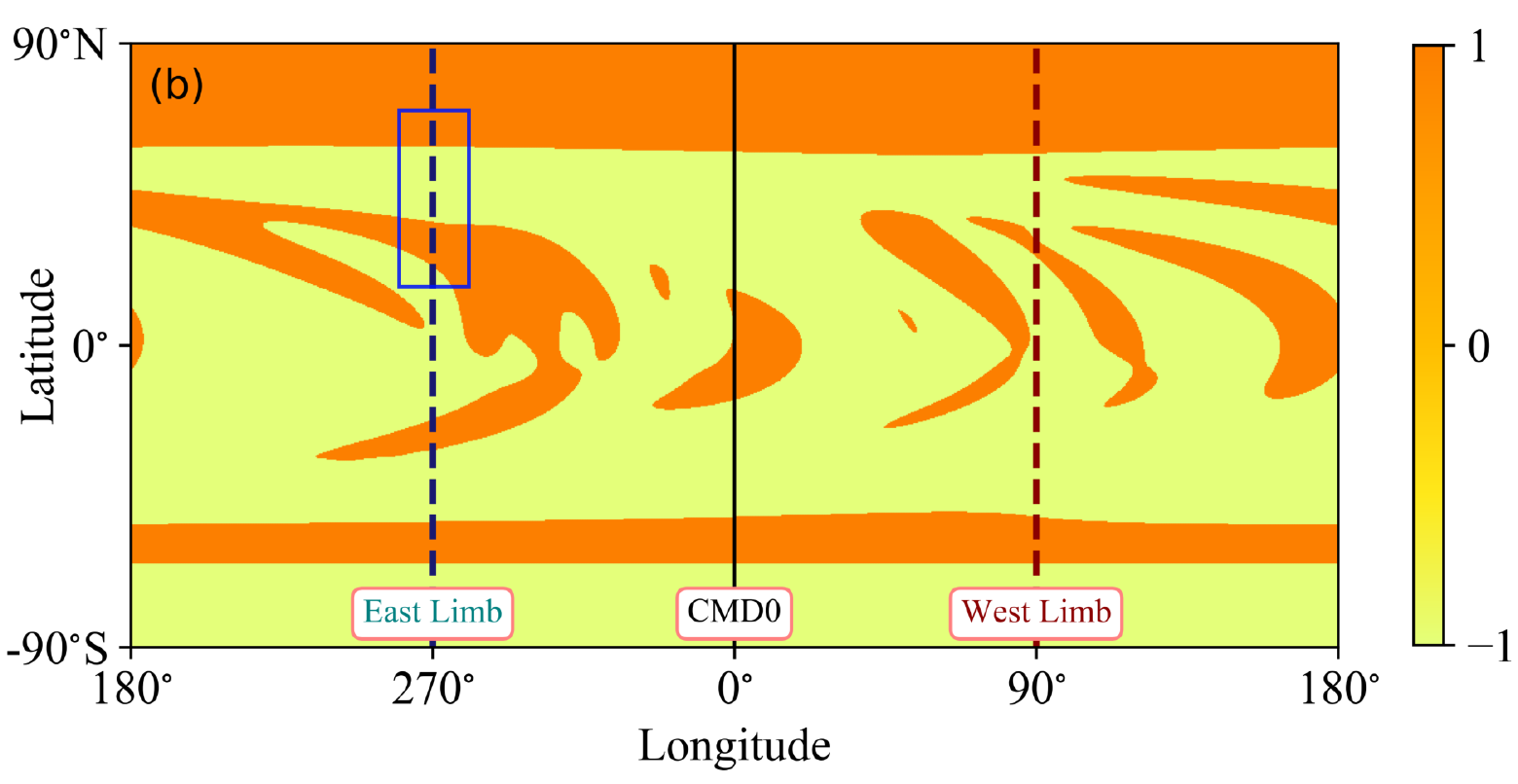}
\caption{(a) Predicted solar surface magnetic field ($B_r$) for 2019 July 2 obtained from the PSSFT model. The colorbar represents the field strength saturated to $\pm 10$ G. (b) Saturated image to indicate the polarity (and inversion lines) of the surface field distribution. The solid black line represents the central meridian, the dashed blue and maroon lines correspond to the east and west limb of the visible solar disk on 2019 July 2. The rectangular box on the east limb represents an area of interest which has the possibility of hosting a pseudo-streamer.}
\label{fig1}
\end{figure}

The PSSFT predicted surface magnetic field distribution corresponding to 2019 July 2 is depicted in Fig.\ref{fig1}(a). The radial component of the magnetic field ($B_r$) obtained from forward running the PSSFT simulation is plotted as a function of latitude and longitude. The region between the east and west limbs represents the surface magnetic field map on the solar disk during the eclipse. As the Sun has entered the minimum phase of solar cycle 24, we witness strong concentrations of unipolar magnetic flux of opposite polarities at both polar caps. This is indicative of a large-scale magnetic configuration with dominant dipolar characteristics. Note that the locally confined clusters of magnetic field near the equatorial region (in this and the following figures) correspond to the residual flux of the emerged active regions and should not be mistaken as new spot emergence. Figure.\ref{fig1}(b) represents the polarity distribution of the surface magnetic field which is evaluated by calculating the quantity, $P(\theta, \phi) = B_r(\theta, \phi)/|B_r(\theta, \phi)|$. This image represents how the magnetic polarity is distributed in the large-scale structures on the surface. 

\begin{figure}
\gridline{\fig{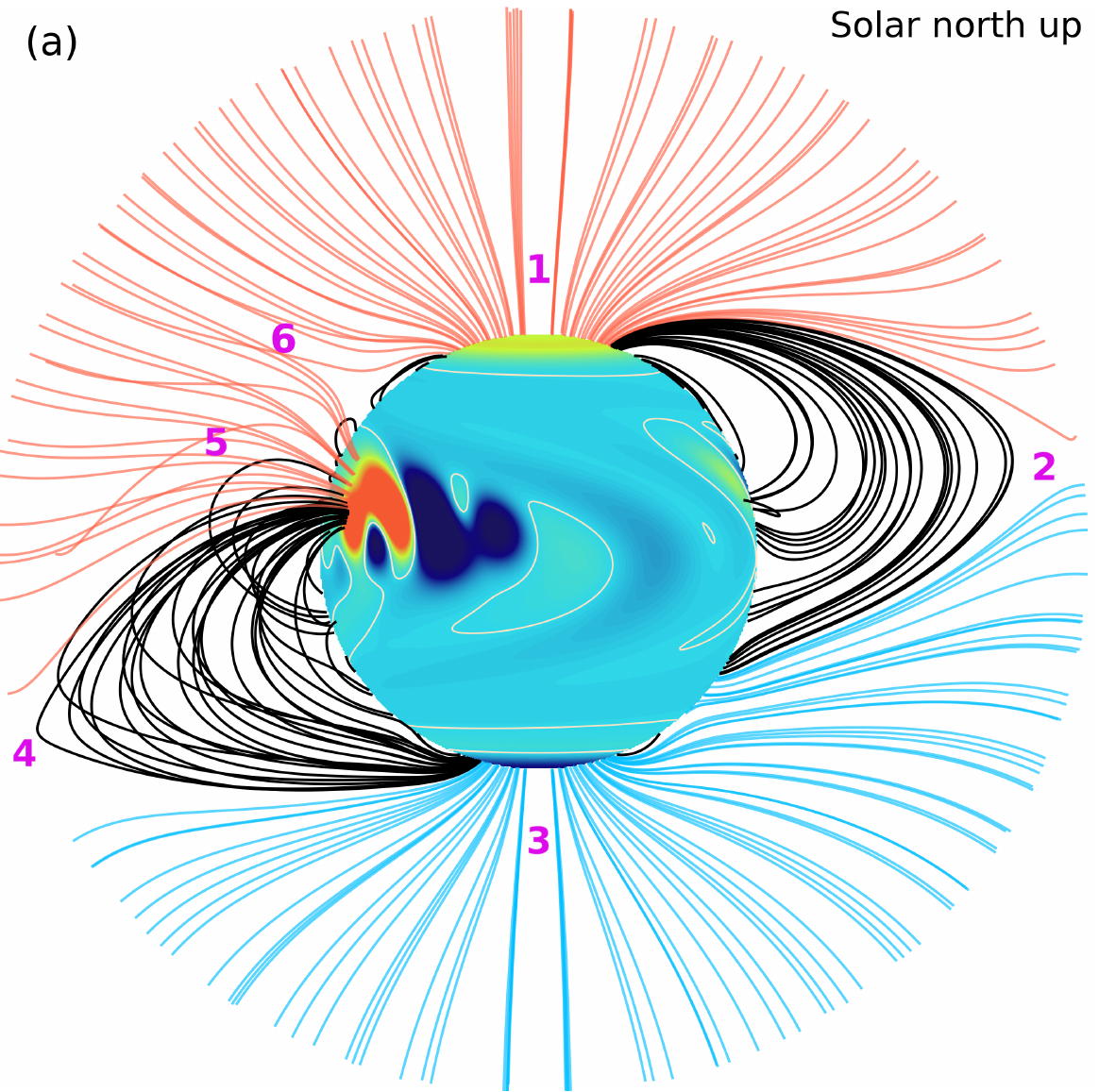}{0.49\textwidth}{}
          \fig{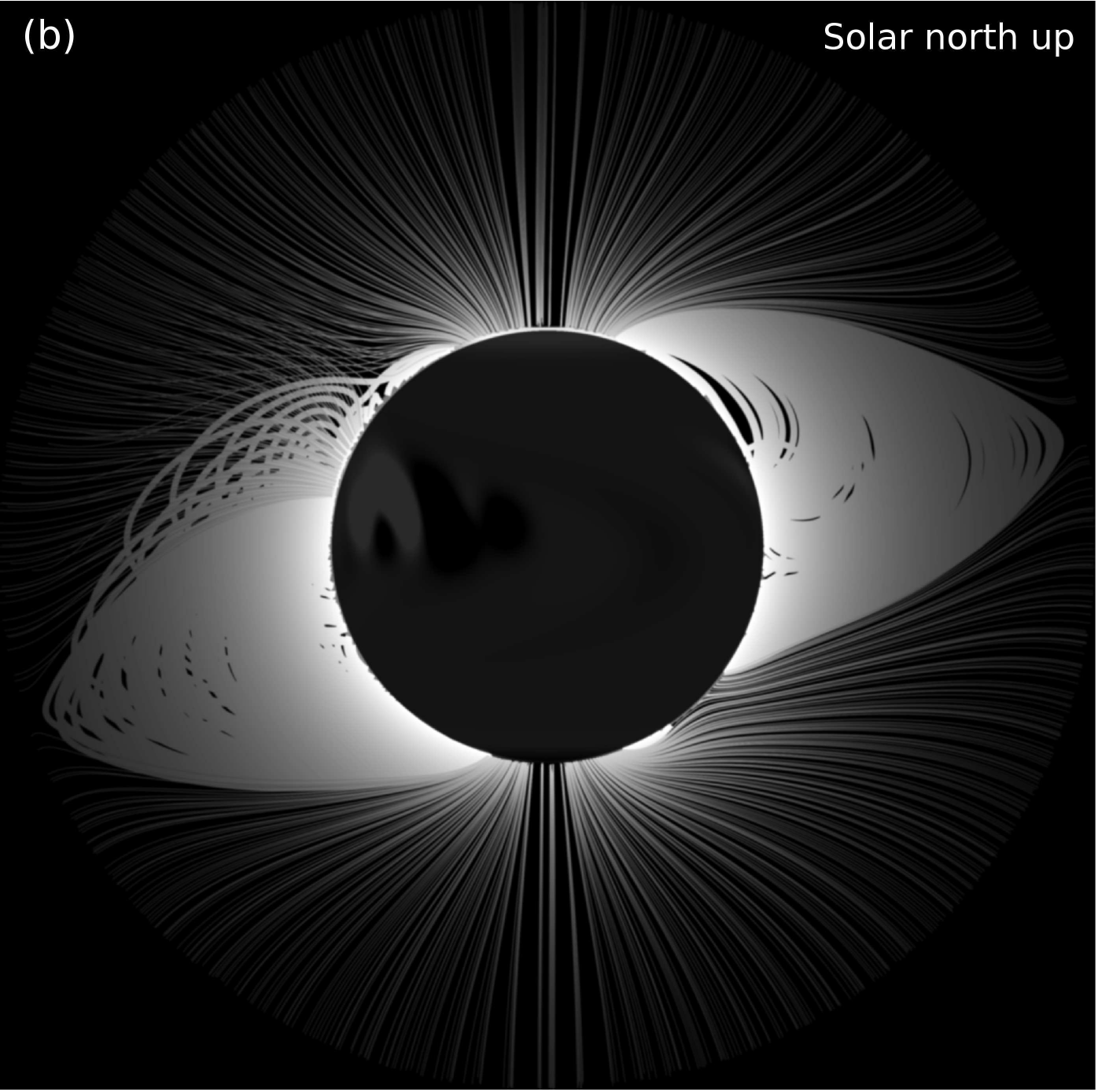}{0.49\textwidth}{}}
\caption{(a) Predicted large-scale coronal structure of the 2019 July 2 solar eclipse with open field lines denoted in  light red (radially outward) and cyan (radially inward). Closed field lines are denoted in black. The white curves on the solar disk correspond to polarity inversion lines. Regions marked $1$ and $3$ denote open field lines (coronal holes) at the polar regions. Regions marked $2$ and $4$ are the cusps of streamers which are predicted to occur on the west (right) and east (left) limbs, respectively, indicating the inclination at which extended coronal plumes may be expected in large-angle coronagraph observations. Region $5$ corresponds to the additional closed field lines adjacent to the helmet streamer on the east limb. Region $6$ overlies a surface field distribution which is symptomatic of a pseudo-streamer host but is unlikely to generate one by 2019 July 2 (see text). (b) A representative white-light corona synthetically constructed from the magnetic field structure to indicate the plausible white light appearance of the corona during the total solar eclipse. }
\label{fig2}
\end{figure}

The predicted surface magnetic field is then utilized in the PFSS extrapolation model as its lower boundary condition to obtain the global coronal magnetic field. The surface field is mapped on the circular disk of Fig.\ref{fig2}(a), where the white curves correspond to the magnetic polarity inversion lines. The PFSS model generated coronal field lines near the limb are extracted to represent the plane of sky corona expected on the day of the eclipse. The locations marked as Regions $1$ and $3$ correspond to the north and the south pole. Based on the polarity of the magnetic field at the foot-points the open field lines are color-coded in light red (radially outward) and cyan (radially inward) while all closed lines are colored in black. The predicted coronal field has two prominent streamer structures, one on the west limb (whose tip or cusp is denoted as Region $2$) and the other on the east limb (Region $4$). The set of closed black curves in Fig.\ref{fig2}(a) corresponding to large-scale closed loop structures separate open flux (coronal holes) of opposite polarities -- the defining characteristic of helmet streamers \citep{wang2007solar}. Although the foot-points of the closed magnetic field lines associated with streamers are extended across the solar equator and major portion of the two hemispheres (north and south) they still have a directional sense. The cusp of the streamer on the east limb (Region $4$) is centred below the solar equatorial plane. The cusp of the streamer on the west limb (Region $2$) is centred above the equatorial plane. Based on this we predict extended coronal plumes visible in large-angle coronagraphs such as LASCO/SOHO to be oriented through the streamer cusps marked as Regions $2$ and $4$. The presence of closed field lines (somewhat inclined to the plane of sky) on the northern edge of the streamer (Region $5$) on the east limb is expected to smooth out the cusp on the northern edge of this streamer and may also host an extended coronal plume beyond the source surface overlying Region $5$. 
 
The large-scale plasma motion or supergranular convection on the photosphere can generate foot-point motion resulting in the rise of the closed loops within the helmet streamers. This can trigger a three-dimensional reconnection with the overlying open field lines. Consequently, plasma materials get energized resulting in heating and emission. In addition, Thomson scattering of photons from regions with enhanced charged particle density (such as in magnetic loops) contribute to the overall appearance of the white light corona (structured by magnetic fields). These are processes we cannot capture with our simplistic coronal field model, however, our coronal magnetic field may be utilized to generate a ``synthetic'' white light corona. Figure.\ref{fig2}(b) is a representation of the white-light corona based on the simulated magnetic field distribution. Here the magnetic field lines are plotted using a single color (white) wherein closed field lines are assigned more weight compared to the open ones. Additionally, an inverse $r^2$ filter is used on the resulting image to generate the ``synthetic'' corona. This image may be taken as a qualitative guide to what the white light corona (within $2.5 R_{\odot}$) might appear like on 2019 July 2, 
 
\subsection{Coronal Magnetometry: Forward Modeled Coronal Polarization Characteristics}

The stokes $I$, $Q$, $U$, $V$ polarization vectors have shown considerable potential to be used as a direct diagnostic of coronal magnetism \citep{lin2004coronal}. The line of sight magnetic field strength provided by the longitudinal Zeeman effect is manifested in circular polarization, $V$. The linear polarization represented by Stokes vectors $Q$ and $U$ originate due to resonance scattering of photons by the electrons in the corona. These Stokes vectors $Q$ and $U$ contain information about the direction of the magnetic field projected onto the plane of sky of the corona. To generate the synthetic Stokes data \citep{judge2006spectral} we utilize the FORWARD tool set \citep{gibson2016forward} using a simple spherically symmetric hydrostatic temperature and density model for the background corona \citep{gibson1999solar}. FORWARD uses the Coronal Line Emission polarimetry code developed by \cite{ judge2001synthesis} to synthesize Stokes ($I$, $Q$, $U$, $V$) line profiles. When the magnetic field is oriented at the Van Vleck angle = 54.74$^{\circ}$ relative to the radial direction (in the solar coordinate system) the linear polarization ($L=\sqrt{Q^2+U^2}$) becomes zero \citep{ van1925quantum}. We refer to the regions where this occur as ``Van Vleck nulls''. The Stokes $V/I$ profile gives the line-of-sight intensity-weighted average magnetic field strength ($B_{LOS}$). 

\begin{figure}[!htb]
\centering
\includegraphics[width=0.40\linewidth]{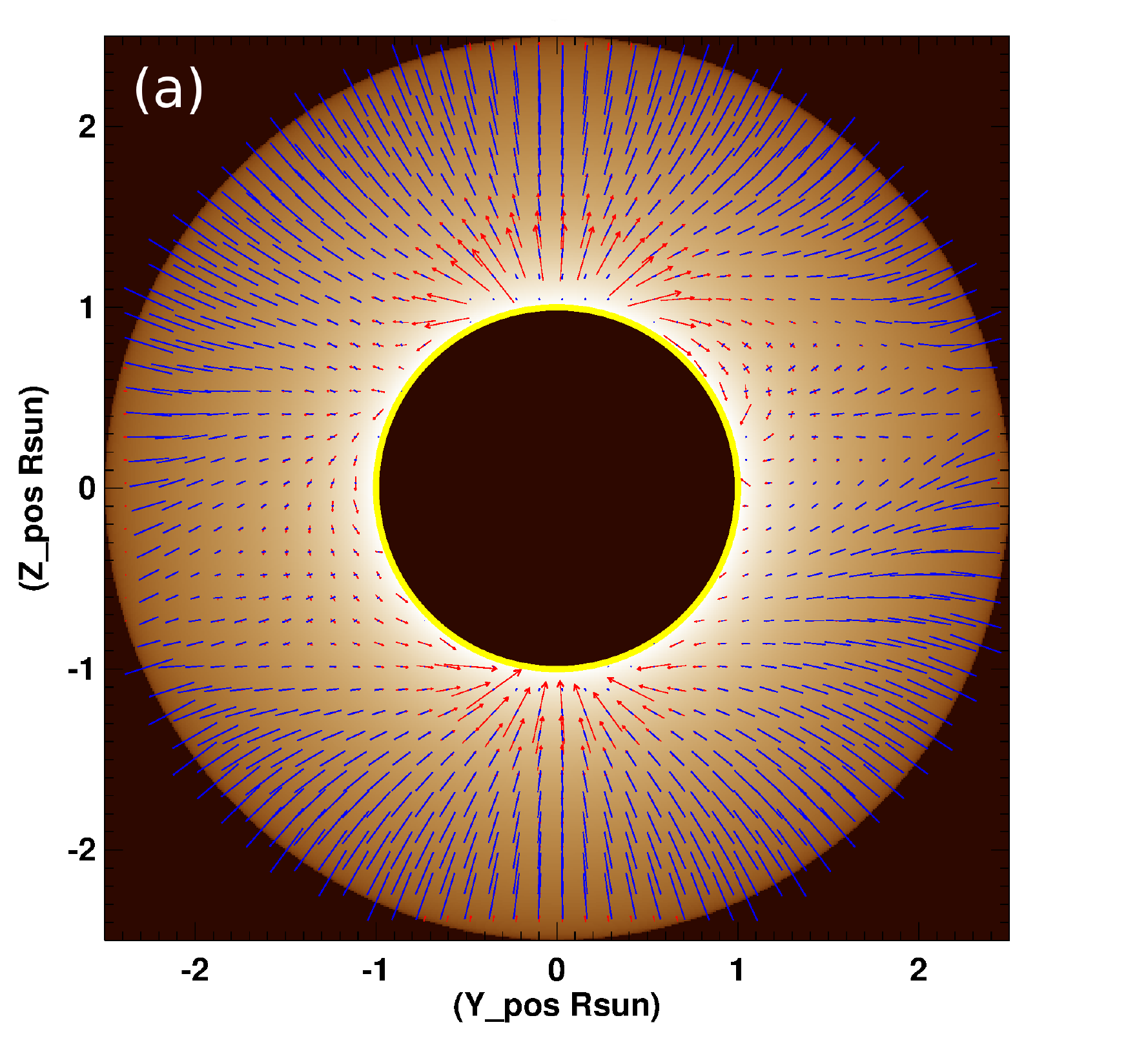}\\
\includegraphics[width=0.40\linewidth]{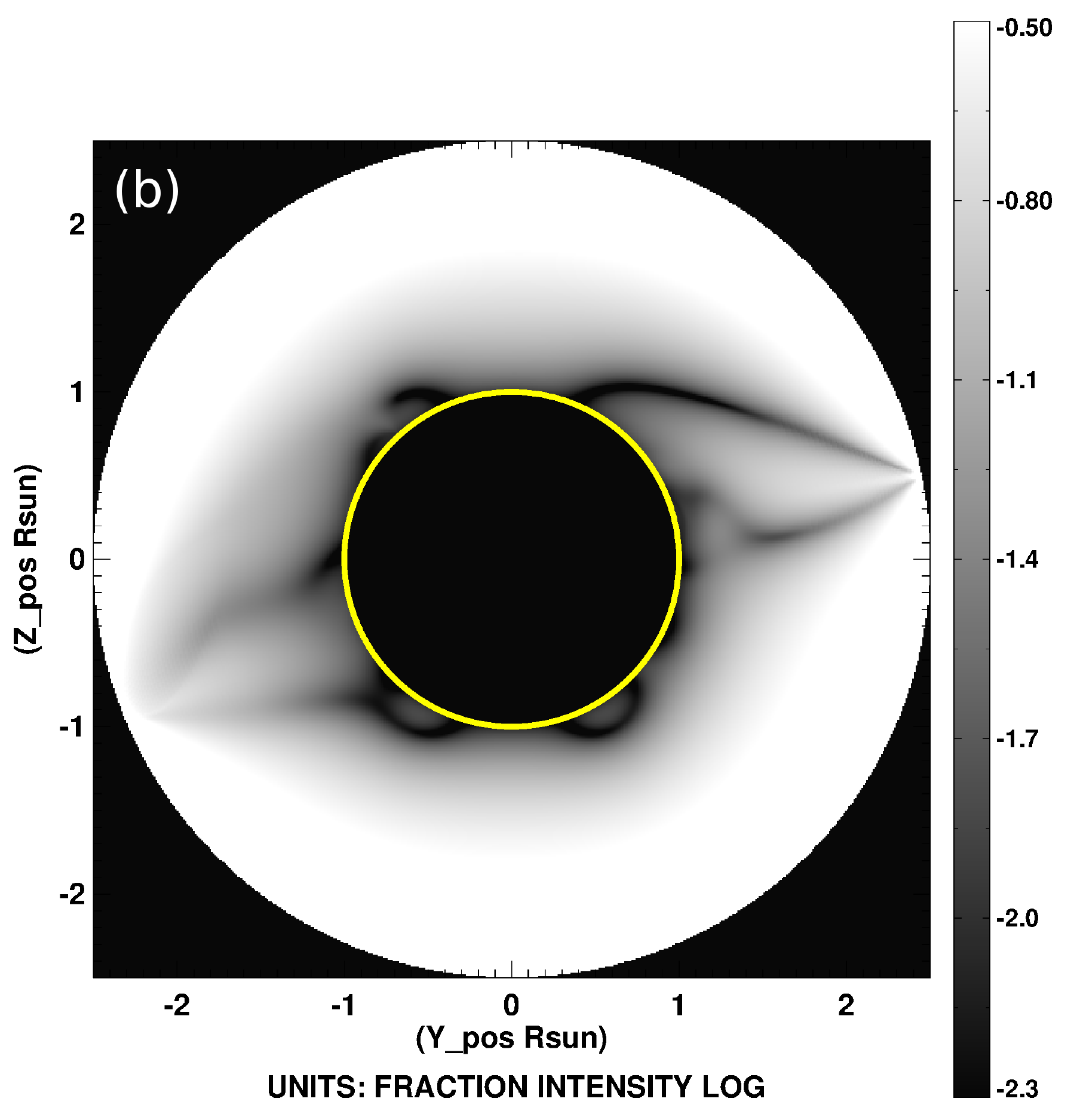}\\
\includegraphics[width=0.40\linewidth]{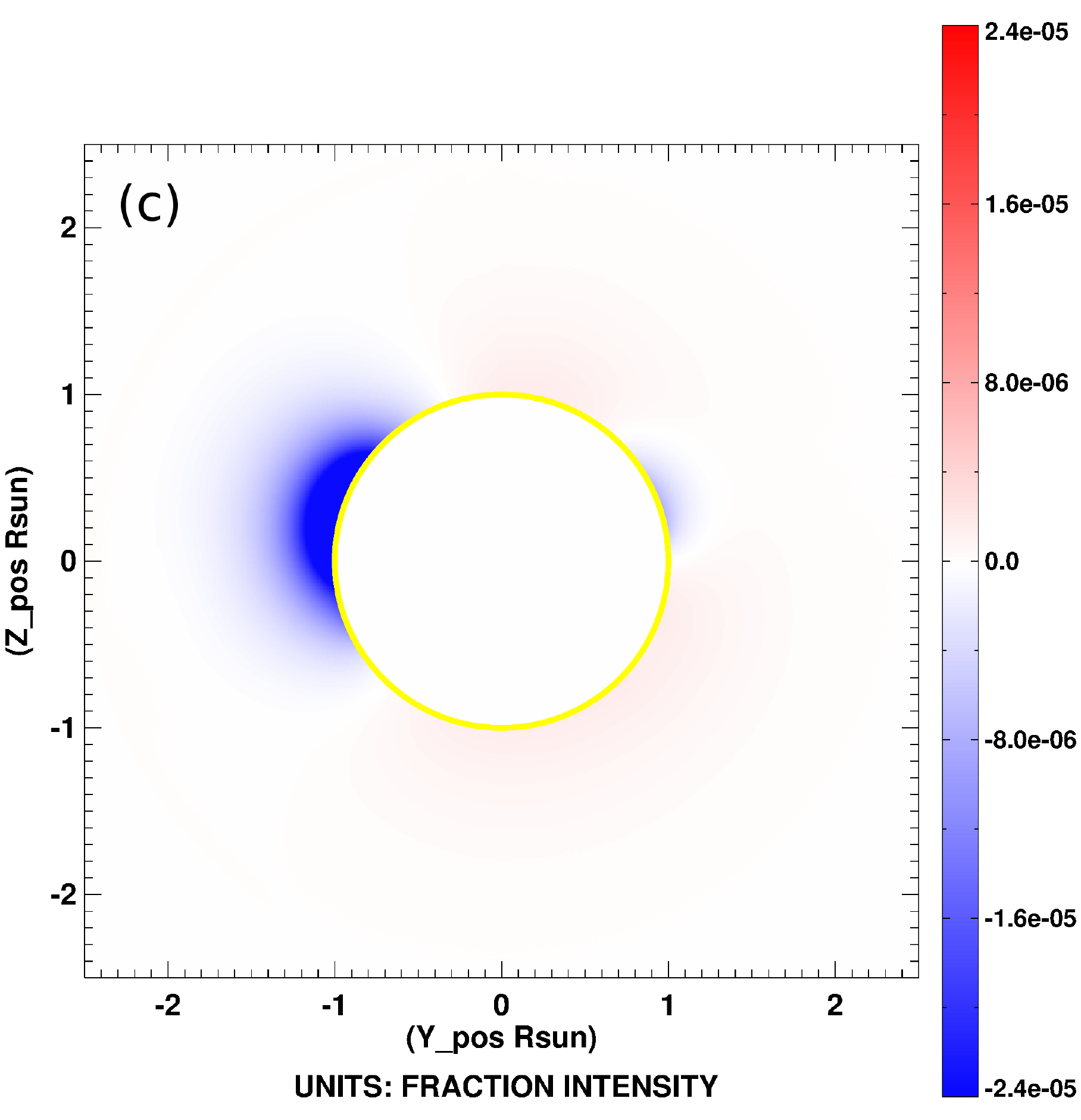}
\caption{Forward modeled coronal polarization maps. (a) Plane of sky magnetic field vectors (red arrows) and linear polarization vectors (blue lines). (b) Degree of linear polarization, $L/I$. (c) The circular polarization given by Stokes $V/I$.}
\label{fig3}
\end{figure}

In Fig.\ref{fig3} we present the FORWARD modelled polarimetric maps derived from the predicted coronal magnetic field structure for 2019 July 2. Figure \ref{fig3}(a) shows the linear polarization vectors (blue lines) corresponding to emission from the Fe XIII transition at 1074.7 nm. The magnetic field vectors in the plane of sky are denoted as red arrows. As expected, the linear polarization vectors correctly identify the direction of the plane-of-sky magnetic field, including the diverging and converging radial fields from the north and south solar poles and the curvature underlying the closed streamer belts. The degree of linear polarization $L/I$ projected on to the plane of sky is shown in Fig.\ref{fig3}(b), wherein, the dark regions corresponding to Van Vleck nulls denote curved field lines of closed streamer loops (oriented at the Van Vleck angle to the local radial direction); note the correspondence with Fig.\ref{fig2}(a). Intriguingly, the low-lying double-loop structure evident at high latitudes in the north-east limb hints at the existence of a pseudo-streamer \citep{Rachmeler2014ApJL}, to which we come back later. Figure.\ref{fig3}(c) depicts the circular polarization given by Stokes $V/I$ which is proportional to the line-of-sight magnetic field strength. Note that Stokes $V \propto B_{LOS} \cos \theta$, where $\theta$ is the polar angle of the magnetic field relative to the line of sight \citep{casini1999spectral}. Therefore, its sign indicates the direction of the field towards or away from the line of sight. Blue shading indicates line of sight integrated magnetic fields directed away from the observer (while red denotes field directed towards the observer). The plot indicates the presence of clock-wise closed loops (as viewed from solar north) connecting the positive polarity patch near the east limb to a negative polarity patch behind the limb which we confirm from the full 3D coronal magnetic field structure (not shown here).     

Such predicted polarization characteristics can, on the one hand, aid in the interpretation of coronal magnetometry studies during solar eclipses. On the other hand, the observations themselves can help constrain coronal field magnetic field models which underlie the forward modeled polarization characteristics.  

\subsection{Evaluating the Possibility of a Pseudo-streamer} 

The coronal magnetic field generated from the PSSFT$-$PFSS coupled model has a narrow collimated structure (marked as Region $6$ on the north-east limb of Fig.\ref{fig2}(a)) with a localized void and very low-lying field loops closing near the surface. Such a magnetic configuration is indicative of a pseudo-streamer which has not quite matured to be visible. Visible pseudo-streamers, in general, materialize under certain conditions in the coronal regions which overlie surface magnetic field distribution where a narrow region of one polarity separates two surrounding opposite polarity open flux patches \citep{wang2007solar,Rachmeler2014ApJL,Abbo2015SoPh}. The surface polarity configuration of this region of interest is highlighted within the rectangular box of the surface magnetic field in Fig.\ref{fig1}(b). We observe a narrow region of negative polarity separating two positive polarity patches on the solar surface and two associated polarity inversion lines. This region is centred around $42^{\circ}$ N on the east limb. Could this region with pseudo-streamer favourable surface magnetic field distribution mature into a visible pseudo-streamer?

A basic pseudo-streamer configuration is characterized by two polarity inversion lines under the cusp of the streamer and a pair of loop arcades (consisting of closed field regions) within the streamer. When conditions are favourable (e.g., induced footpoint motions due to shearing or differential rotation) the closed loops can rise and undergo two types of interchange reconnection with the surrounding open field lines. One possibility is that closed field lines at the outer edge of any arcade can undergo three dimensional reconnection with adjacent open field lines. Another possibility is that sheared closed field lines can rise and expand in to the corona where they reconnect with opposite polarity open field lines across the null point (near the source surface). The former necessitates loop arcades with field strength similar to the adjacent open field and the latter necessitates loop arcades which are high enough to form an x-point with the adjacent open field lines.  

\begin{figure}
\gridline{\fig{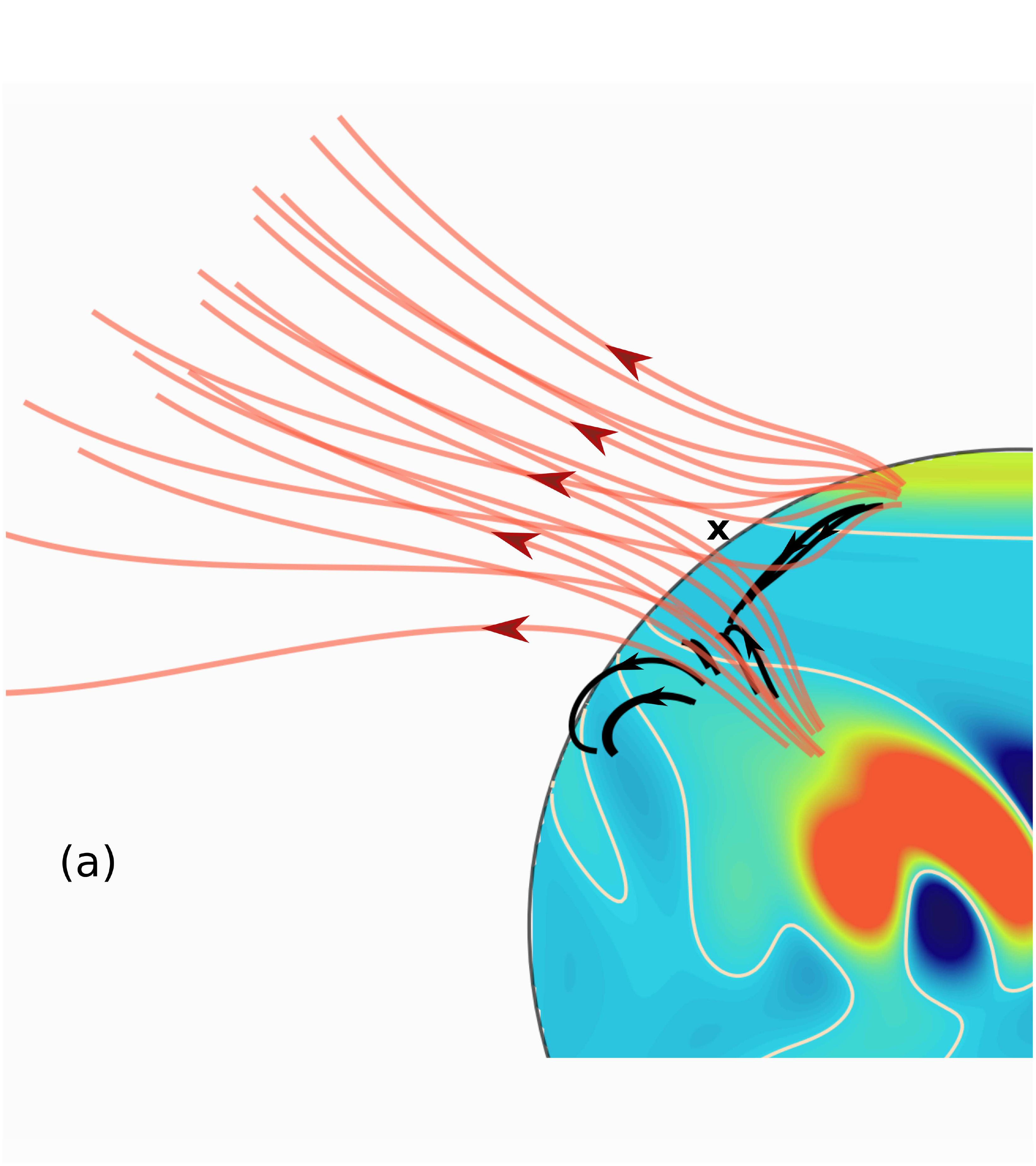}{0.49\textwidth}{}
          \fig{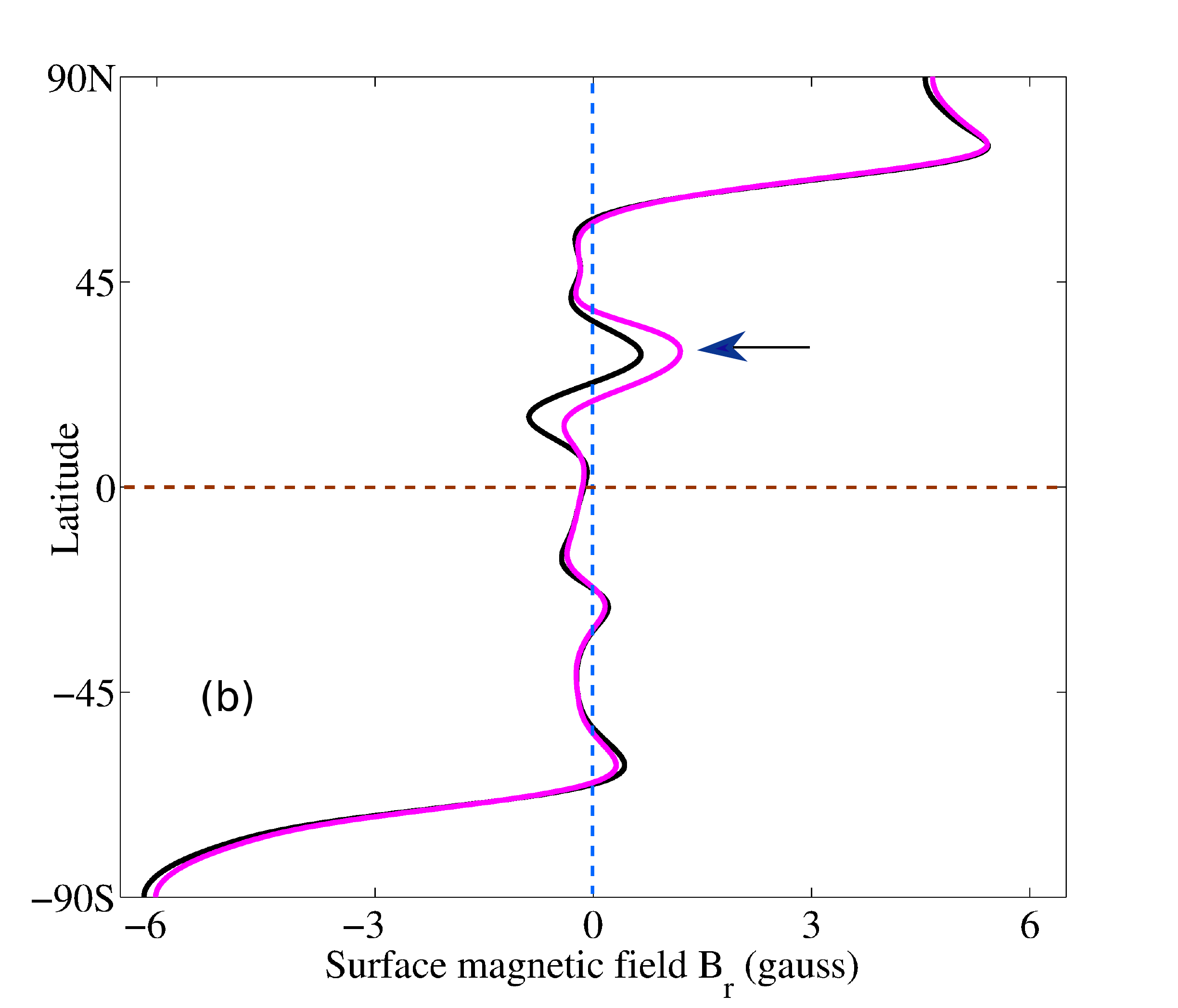}{0.49\textwidth}{}}
\caption{(a) The predicted coronal magnetic field line distribution around ``Region $6$'' (with reference to Fig.\ref{fig2}). The weak field closed loops denoted in black are extremely low-lying and enclosed by open field lines denoted in light red. (b) The black curve represents the surface magnetic field strength along the east (left) limb on the day of the eclipse. The magenta curve denotes the PSSFT model predicted field strength along the same limb one solar rotation later indicating the possibility of increased positive radial field transport to the edge of the region of interest.}
\label{fig4}
\end{figure}

We investigate the coronal magnetic field configuration and the underlying surface magnetic field distribution in this region of interest in Fig.\ref{fig4}. We find that the weak field closed loops (black) are extremely low-lying and have not yet matured in to arcade-like structures (see Fig.\ref{fig4}(a)). Given the weak underlying fields at this time they are also unlikely to reach higher altitudes to form an x-point with the open field lines (light red). The possibility of interchange reconnection that could trigger a visibly stable pseudo-streamer by 2019 July 2, is therefore, low. Intriguingly, however, a longer forward run of the PSSFT model for a further solar rotational timescale indicates that transport of positive radial flux towards the north pole is still ongoing which is expected to increase the field strength of the open field patch just south of the region of interest; note the increasing height of the local peak indicated by the arrow in Fig.\ref{fig4}(b). We cannot at this time rule out the possibility that this future flux pile-up could give rise to a visible pseudo-streamer under favourable conditions that may occur after the eclipse of 2019 July 2.

\section{Concluding Discussions}

In summary, here we predict the Sun's coronal magnetic field structure expected to be observed during the total solar eclipse of 2019 July 2. We also present forward modeled polarization characteristics of the coronal magnetic field that should inform and aid in the interpretation of observations during the eclipse.

It is our belief that the usage of the PSSFT model allows for better prediction of the surface magnetic field, especially at high latitudes enabled via surface plasma flux transport processes. Such models assimilating surface magnetic field data provide ideal boundary conditions for simulating and predicting the large-scale coronal structure. We note that our scheme is not perfect. We have ignored any non-linear effects on the plasma transport in the surface flux evolution (whose impact is expected to be minimal during declining phases of the cycle). Also PFSS extrapolations for simulating the corona cannot account for current carrying large-scale sheared structures. Neither can we account for the impact of the heliospheric current sheet near the source surface and beyond, interactions with which can deflect the cusps of streamer belts and coronal plumes at larger distances from the Sun. Detailed comparisons with observations are expected to pinpoint the deficiencies which need to be addressed. Based on such comparisons we plan to refine our predictive scheme with the addition of more advanced MHD models which can capture a broader range of coronal phenomena more accurately.  

The coronal magnetic fields, which evolve in response to driving from the solar surface, govern spatial and temporal variations of the slow and fast components of the solar wind and heliospehric open flux. They also spawn solar flares and CMEs which have severe space weather impacts. Several future facilities are focusing on coronal magnetometry and gearing up to return coronal magnetic field measurements. These include the ground-based facilities DKIST\footnote{\url{https://en.wikipedia.org/wiki/Daniel_K._Inouye_Solar_Telescope}} and CoMP \citep{CoMP2008SoPh}. ISRO's Aditya-L1 space mission, currently under development, will fly the Solar Ultraviolet Imaging Telescope (SUIT; \citealt{SUIT2017cursci}) and the Variable Emission Line Coronagraph (VELC; \citealt{VELC2017cursci}) instruments which would simultaneously observe filament-prominence-arcade systems and coronal magnetic fields. We expect these multi-viewpoint, multi-wavelength observations to revolutionize the field of coronal magnetometry. Collective endeavours combining theoretical modeling with these coronal observations are expected to lead to refined data-driven operational forecasting models for solar activity induced space weather.

\acknowledgements CESSI is funded by the Ministry of Human Resource Development, Government of India. S.D. acknowledges funding from the INSPIRE program of the Department of Science and Technology,  Government of India. We acknowledge utilization of data from the NASA/SDO HMI instrument maintained by the HMI team and the Royal Greenwich Observatory/USAF-NOAA active region database compiled by David H. Hathaway. Additional graphics and text related to this prediction are available online at \url{http://www.cessi.in/solareclipse2019}.

\bibliographystyle{aasjournal}

\end{document}